\documentclass[%
preprint,
showpacs,
 bibnotes,
 amsmath,amssymb,
 aps,
prb,
floatfix,
]{revtex4-1}

\usepackage{graphicx}
\usepackage{dcolumn}
\usepackage{bm}

\usepackage{float}
\usepackage{hyperref}

\usepackage[usenames]{xcolor}

\usepackage{multirow}

\begin{document}
\newfloat{scheme}{htbp}{lof}
\floatname{scheme}{\small SCH.~}

\preprint{APS/123-QED}

\title{\textit{Ab initio} calculations of the concentration dependent band gap reduction in dilute nitrides}

\author{Phil Rosenow$^1$}
\author{Lars C. Bannow$^2$}
\author{Eric W. Fischer$^1$}
\author{Wolfgang Stolz$^3$}
\author{Kerstin Volz$^{2,3}$}
\author{Stephan W. Koch$^{2,3}$}
\author{Ralf Tonner$^{1,3}$}%
\email{tonner@chemie.uni-marburg.de}
\affiliation{%
 $^1$Fachbereich Chemie, Philipps-Universit\"at Marburg, Hans-Meerwein-Stra\ss e 4, 35032 Marburg, Germany
}%
\affiliation{%
 $^2$Fachbereich Physik, Philipps-Universit\"at Marburg, Renthof 5, 35032 Marburg, Germany
}%
\affiliation{
 $^3$Material Sciences Center, Philipps-Universit\"at Marburg, Hans-Meerwein-Stra\ss e 6, 35032 Marburg, Germany
}%

\date{\today}

\begin{abstract}
While being of persistent interest for the integration of lattice-matched laser devices with silicon circuits, the electronic structure of dilute nitride III/V-semiconductors has presented a challenge to \textit{ab initio} computational approaches.
The root of this lies in the strong distortion N atoms exert on most host materials.
Here, we resolve these issues by combining density functional theory calculations based on the meta-GGA functional presented by Tran and Blaha (TB09) with a supercell approach for the dilute nitride Ga(NAs).
Exploring the requirements posed to supercells, we show that the distortion field of a single N atom must be allowed to decrease so far, that it does not overlap with its periodic images.
This also prevents spurious electronic interactions between translational symmetric atoms, allowing to compute band gaps in very good agreement with experimentally derived reference values. These results open up the field of dilute nitride compound semiconductors to predictive \textit{ab initio} calculations.


\end{abstract}


\maketitle


\section{Introduction}
In recent years, semiconductor compounds based on gallium arsenide (GaAs) containing small amounts of nitrogen (N) have proven to be promising materials for all kinds of optoelectronic applications. One of the first applications were (GaIn)(NAs) laser diodes. \cite{Kondow1996} More recently, silicon-matched Ga(NAsP) structures as laser active materials have become a highly active area of research. \cite{Kunert2006} Since the band gap of GaN was known to be larger than that of GaAs, a blueshift of the photoluminescence (PL) edge was expected for Ga(NAs). However, the first experimental PL of dilute nitride Ga(NAs) measured by \textcite{Weyers1992} showed a redshift of the PL. Further studies showed, that the underlying band gap bowing was unusually strong. \cite{Kondow1994,Uesugi1999} In contrast to other III/V-semiconductors, the description of bowing within the commonly used virtual crystal approximation agrees less well with experimental band gaps than the slightly more involved conduction band anti-crossing (CBAC) model. \cite{Vurgaftman2003} While this dependence allows to specifically manipulate the band gap by controlling the composition, the unexpected strength of this effect also called for an explanation.

Based on density functional theory (DFT) and quasiparticle (GW) calculations, \textcite{Rubio1995} were able to attribute this effect to an increase of the volume caused by the lattice mismatch, lowering the energy of the conduction band (CB) edge, which is strongly localized on the N atoms.
Further studies, mainly based on DFT within the local density approximation (LDA) were performed in the following years. \cite{Wei1996,Bellaiche1996,Bellaiche1997,Zhang2003}
A study based on the hybrid functional HSE06 showed an improved agreement with the observed band gap bowing. \cite{Virkkala2012}
Regarding the origin of the strong bowing, further studies found that the CB states centered on the N atoms are extended especially along the zigzag lines in the zinc blende structure. \cite{Kent2001,Virkkala2013}

In the context of our work, it is noteworthy that none of the cited \textit{ab initio} results reproduced the band gaps accurately.
To correct this, Zunger \textit{et al.} constructed empirical pseudopotentials fitted to GW band structures, experimentally determined band gaps and LDA deformation potentials.
This allowed them to perform calculations with huge supercells (13824 atoms). \cite{Bellaiche1996,Bellaiche1997,Kent2001}
Thereby, they showed that the large lattice mismatch between As and N leads to long range strain fields. \cite{Kent2001} This, and the extension of N-centered CB states indicates the need for large supercells. \cite{Zunger1999}

In order to further the development of optoelectronic devices based on dilute nitrides, it is desirable to predict the optical properties of new materials.
This requires an accurate \textit{ab initio} description of the band structure and the band gap without using empirical or experimental parameters.
While DFT is most often the method of choice in material science, there are two factors that usually prevent a quantitative description using DFT supercell calculations:
Firstly, commonly used density functionals mostly underestimate the band gap of semiconductors considerably (LDA,\cite{Perdew1992,Perdew1985} PBE\cite{PBE}).
Hybrid functionals including exact exchange like HSE06\cite{Krukau2006} (more recently used with tailored parameters\cite{Moussa2012}) can solve this issue but are computationally more demanding and hardly applicable to supercells with several hundred atoms in a routine fashion.
Secondly, the enormous lattice mismatch between GaN and GaAs due to the size difference of As and N (the lattice constant of GaN is roughly $80\,\%$ that of GaAs) in combination with the periodicity of the finite size supercells cause an artificial reduction of the band gap. \cite{Kent2001,Virkkala2013}
Overcoming the latter problem requires large supercells which adds to the former by increasing the computational cost.
An alternative approach is to use special quasirandom structures (SQS) as proposed by Zunger \textit{et al.}, which aim to emulate a random distribution in a semiconductor alloy.\cite{Zunger1990} This approach has often been successfully used in the past but requires convergence with respect to cell size nonetheless.
Therefore, a functional capable of accurately describing the band gap while still being computationally efficient is needed.

The recently developed meta-GGA functional TB09 (also known as mBJLDA or TB-mBJ)\cite{Tran2009} is a promising candidate to fulfill these requirements for Ga(NAs).
It has been shown that TB09 allows to obtain band gaps commonly in good agreement with experimental data\cite{Tran2009,Kim_PRB_82_2010,Jiang_JCP_138_2013} while being computationally very efficient in comparison to HSE06 and GW calculations. \cite{Tran2009}

Thus, in this work we will use the TB09 functional in combination with fairly large supercells (up to 432 atoms) to solve the long standing issue of the \textit{ab initio }prediction of accurate band gaps in dilute nitrides. In Section \ref{Sec:methods} we introduce our computational methods and parameters.
Thereafter, we present our results, starting with a brief validation of the approach used followed by a discussion of the structural effects of nitrogen incorporation and finally the effect on the band gap for different supercells.

\section{Computational Method\label{Sec:methods}}
DFT calculations were performed using the Vienna \textit{ab initio} Simulation Package (VASP 5.3.5)\cite{VASP1993, VASP1994, VASP1996a, VASP1996b} with a plane wave basis in conjunction with the projector-augmented wave method. \cite{Blochl1994, Kresse1999}
The basis set energy cut-off was set to $450\,\text{eV}$ for optimizations and $350\,\text{eV}$ for large supercell band gap computations.
The reciprocal space was sampled with a $\Gamma$-centered Monkhorst-Pack grid with six intersections per direction for primitive unit cells and an accordingly reduced set for supercells. \cite{monkhorstpack}
Cell relaxations of primitve cells and all lattice relaxations were done with the PBE functional, \cite{PBE} corrected for dispersion interactions with the DFT-D3 scheme. \cite{GrimmeD3, GrimmeD3BJ}
Lattice parameters for binary materials were derived from fitting the Vinet equation of state, yielding a theoretically optimized value.\cite{Vinet1986}
The lattice parameters for ternary cells were linearly interpolated from the constituents (Vegard's rule).
For $6^3$ supercells with more than one N atom, sampling was performed by randomly selecting As atoms to be replaced by N atoms and averaging over few arrangements.
For a small number of N atoms, more qualitatively distinct arrangements (e.g. different average distances) are possible; thus, more arrangements were sampled for smaller concentrations.
Calculations of band gaps and band structures used the TB09 functional\cite{Tran2009} as well as PBE0\cite{Perdew1996}, HSE06\cite{Krukau2006}, and LDA-1/2\cite{Ferreira2008,Ferreira2011} functionals for comparison.
Potentials for the LDA-1/2 calculations were prepared using the ATOM code with a cut-off (in a.u.) of 0.9875 for Ga, 3.7725 for As, 3.6550 for P, and 2.9275 for N. \cite{Ferreira2008,atomico}
In addition, G$_0$W$_0$ calculations were carried out based on the PBE orbitals.
For all band gap computations, spin-orbit coupling was considered, with the exception of G$_0$W$_0$. For relaxations, the electronic energy and forces were converged to $10^{-6}\,\text{eV}$ and $10^{-2}\,\text{eV}\cdot \text{\AA}^{-1}$, respectively. The electronic energy and eigenvalues for band gap computations in supercells were converged to $10^{-4}\,\text{eV}$.

A supercell calculation yields an energy dispersion $E(\mathbf{K})$ where $\mathbf{K}$ is the wave vector in the reciprocal space of the supercell.
Often, it is desirable to project the supercell eigenstates $|\mathbf{K},n\rangle$ on the eigenstates $|\mathbf{k}_j,m\rangle$ of the respective primitive cell.
The index $j$ accounts for the fact that every supercell eigenstate matches $N^3$ primitive cell eigenstates where $N$ depends on the size of the supercell.
Thus, a so called \textit{effective} band structure (EBS) $E(\mathbf{k})$ can be extracted.
For this, we calculate the spectral weight that is a measure for the Bloch character of a specific eigenstate.
We follow the steps outlined in the appendix of Ref.~\cite{Popescu2012}.
The spectral weight $w_{n,\mathbf{K}}(\mathbf{k})$ is given by the square sum of the relevant plane wave coefficients
\begin{equation}
w_{n,\mathbf{K}}(\mathbf{k}_j) =
\sum_{\mathbf{g}}|C_{n,\mathbf{K}}(\mathbf{g}+\mathbf{G}_j)|^2
\end{equation}
where $n$ is the band index and $\mathbf{g}$ and $\mathbf{G}_j$ are reciprocal lattice vectors of the primitive cell and the supercell, respectively.

Apart from cubic supercells, SQS cells have been used for ternary cells. \cite{Zunger1990} SQS cells have been generated with the Alloy-Theoretic Automated Toolkit (ATAT), \cite{VanDeWalle2013} with correlations between two atoms up to the third sphere of the same lattice site, three atoms up to the second sphere and four atoms in the first sphere only, corresponding to four distinct two-atom clusters, ten three-atom clusters and two four-atom clusters.

The $(N\times N \times N)$ supercells of the primitive zinc blende unit cell will be referred to as $N^3$ supercells, SQS cells with $N$ atoms as SQS-$N$.
Structures and band gaps for the used cells are shown in the supporting information. 

\section{Results and Discussion}

\subsection{Method benchmark}
First, we revisit the question of band gap computation for common functional classes.
TABLE~\ref{tab:functionals} shows a comparison of band gaps from PBE, LDA-1/2, PBE0, HSE06, and TB09 together with the quasiparticle approach G$_0$W$_0$ for binary III/V-semiconductors GaAs, GaN and GaP at theoretically optimized lattice constants.
While PBE shows the well-known shortcoming of GGA functionals for band gap computations, the LDA-1/2 approach and the hybrid functionals achieve a decent agreement both with the many-body approach and experimental values.
However, the TB09 functional clearly outperforms the other density functionals and competes very well with the much more demanding GW-method, indicating its suitability for the band gap calculation of large supercells. Thus, we selected this functional to compute the electronic structure of the compounds.

\begin{table}
\caption{Direct band gap (eV) of III/V-semiconductors in zinc blende structure with various methods. G$_0$W$_0$ is based on PBE and without spin-orbit coupling. Lattice constants (GaAs: 5.689~\AA, GaN: 4.580~\AA, GaP: 5.477~\AA) were theoretically optimized (PBE-D3) as described in the method section. The root mean square deviation (RMSD) w.r.t. experimental reference values is given for each method.}\label{tab:functionals}
\begin{ruledtabular}
\begin{tabular}{lccccccc}
 & PBE & LDA-1/2 & PBE0 & HSE06 & TB09 & G$_0$W$_0$ & Ref.\cite{Vurgaftman2001} \\
 &     &         &      &       &      & (PBE)      & \\
\hline
GaAs & 0.32 & 1.17 & 1.71 & 1.11 & 1.44 & 1.41 & 1.52 \\
GaN  & 1.58 & 3.16 & 3.48 & 2.78 & 3.03 & 3.09 & 3.28 \\
GaP  & 1.74 & 2.58 & 3.31 & 2.66 & 2.95 & 2.86 & 2.86 \\
RMSD & 1.36 & 0.27 & 0.30 & 0.39 & 0.16 & 0.13 & \\
\end{tabular}
\end{ruledtabular}
\end{table}

\subsection{Structural effect of N-incorporation compared to other group V elements}
\begin{figure}
 \includegraphics[width=.6\textwidth]{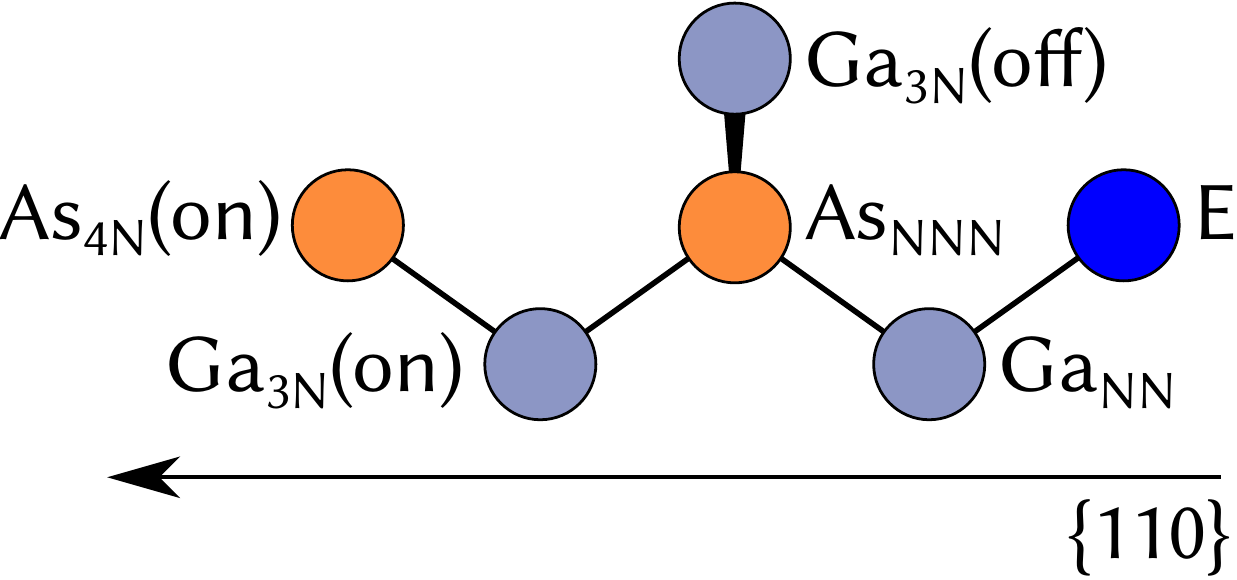}
 \caption{Schematic representation of the environment of an atom E in GaAs lattice with E=N, Bi.}\label{fig:positionscheme}
 \end{figure}

It has been shown before that from the available group V atoms in the GaAs lattice, incorporation of N results in the by far strongest displacement of nearest neighbor Ga atoms. The N incorporation leads to roughly twice the displacement from ideal lattice position ($-0.379$\AA) compared to the other extreme in that group, Bi ($+0.190$~\AA).
This lattice distortion enhances the extension of the nitrogen state in real space, causing the interaction with translational images of the atom. This is the underlying cause for the artificial band gap reduction described above.
Beyond nearest neighbors, the distortion propagates further and is highly anisotropic.
We show this for a $5^3$ supercell with one As-atom replaced by N and Bi, respectively.
Both cells have the lattice constant of the respective compound and the atomic positions have been relaxed.
The distortion is shown in TABLE~\ref{tab:distortions} in terms of displacement for an atom at a given position relative to atom E (E = N, Bi).
The labeling of these positions is explained in FIG.~\ref{fig:positionscheme} and is based on the connectivity to atom E along chemical bonds rather than spatial distance.
For the first two coordination spheres, i.e. the nearest neighbors (NN) and next-nearest neighbors (NNN), the distortion is isotropic.
At the third-nearest neighbors (3N) positions, a distinction can be made between the Ga-atoms along the \{110\} zig-zag chains ($\mathrm{Ga_{3N}(on)}$) and those sitting off these chains ($\mathrm{Ga_{3N}(off)}$).
While the former are further away from atoms E in terms of spatial distance than the latter, their displacement is about four times as large.
Along the \{110\}-directions, every step beyond the next-nearest neighbor spheres approximately halves the distortion.
Comparing Ga(NAs) and Ga(AsBi), the distance over which the GaAs lattice surrounding atom E is relaxed is twice as large for the dilute nitride as for the bismide, in accordance with their difference in atomic radius.
This explains, why dilute bismides can be described well with moderately sized supercells\cite{Bannow2016} in contrast to dilute nitrides.

\begin{table}
\caption{Displacement relative to ideal lattice site of selected atoms in Ga$_{125}$(As$_{124}$N$_1$) compared to Ga$_{125}$(As$_{124}$Bi$_1$) in \AA. Labeling of position relative to atom E and according to FIG.~\ref{fig:positionscheme}}\label{tab:distortions}
\begin{ruledtabular}
\begin{tabular}{lcc}
    & Ga$_{125}$(As$_{124}$N$_1$) & Ga$_{125}$(As$_{124}$Bi$_1$) \\
\hline
Ga$_\text{NN}$      & $-0.379$ & $0.190$ \\
As$_\text{NNN}$     & $-0.110$ & $0.053$ \\
Ga$_\text{3N}$(on)  & $-0.067$ & $0.036$ \\
Ga$_\text{3N}$(off) & $-0.017$ & $0.008$ \\
As$_\text{4N}$(on)  & $-0.032$ & $0.019$ \\
Ga$_\text{5N}$(on)  & $-0.005$ & $0.003$ \\
\end{tabular}
\end{ruledtabular}
\end{table}

Since the displacement is halved with every step to the next position, it should decay to sub-pm magnitude with the 6N-position in the labeling used here.
In the $5^3$ supercell, the 5N-position is already halfway between the N atom and its periodic image and is as such balanced by those (the displacement is perpendicular to the connection between the N-atom and its periodic image and a result of the displacement of neighboring atoms).
By extrapolation to the next largest cell size, the strain field can be expected to be sufficiently decayed.

\subsection{Band gap evolution for small and large supercells}
In order to study the effect of the cell size on the band gap of a dilute nitride, it is instructive to compare properties of two cubic supercells with the same N-concentration.
Thus, two supercells of different size for $3.7\,\%$ N have been taken to study different contributions to the band gap reduction due to N incorporation.
The smaller one is a $3^3$-supercell (54 atoms in total, one N), the larger one a $6^3$-supercell (432 atoms in total, eight N).
For the larger supercell, four different random arrangements of the N atoms have been sampled (see Section \ref{Sec:methods}).
One of these showed excellent agreement with the CBAC reference and will be used as an example in the remainder of this paragraph (see FIG.~\ref{fig:ganas-chgdens}b for the structure).
Following \textcite{Kent2001}, different contributions to the band gap change can be distinguished.
In the first step, the lattice constant of GaAs is changed to that of the Ga(NAs) compound, which increases the band gap to $1.61\,\text{eV}$, see TABLE~\ref{tab:gap-evolution}.
An appropriate number of As atoms in each cell is replaced by N atoms in the second step, without relaxing the structure.
This changes the band gap by different amounts for each supercell size: by $-0.42\,\text{eV}$ for the $3^3$ supercell and by $-0.22\,\text{eV}$ for the $6^3$ supercell, showing the smaller finite size effect for the latter.
In the third step, the lattice positions are relaxed, further reducing the band gap to $0.58\,\text{eV}$ for the $3^3$ and to $1.06\,\text{eV}$ for the $6^3$ supercell.
Especially this last step clearly shows the spurious interactions of translational images in small supercells.

Finally, it is instructive to replace the N atoms in the relaxed structure by As atoms again, in order to separate electronic and structural effects.
In this case, the band gap of the $3^3$ supercell is $0.12\,\text{eV}$ smaller than that of the $6^3$ supercell, which emphasizes the dampened effect of the strain field in the latter.
It should be stressed, that these contributions are not additive in either case and a synergy between both effects can be observed.
Both the single and combined effect is smaller for the larger supercell.

The relaxation step in the procedure described above can be further divided in two different ways.
First, following the band gap along the relaxation by computing it for partly relaxed structures, it can be seen that the band gap reduction is linear with relaxation, showing a simple relationship between structural deformation and change in band gap.
This emphasizes the stronger effect on the band gap when combining strongly mismatched materials.
Secondly, relaxing only the four nearest neighbors of the N atom accounts for $97\,\%$ of the band gap reduction, showing that the overlap between N and nearest neighbor-Ga atoms is most influential for the stabilization of the states at the CB edge.

\begin{table}
\caption{Band gaps of the $3^3$ and one $6^3$ supercells with $3.7\,\%$ N for the unrelaxed and relaxed structure and the relaxed structure, where N has been replaced with As. All values in eV.}\label{tab:gap-evolution}
\begin{ruledtabular}
\begin{tabular}{cccccc}
Supercell & GaAs & GaAs($a_\text{Ga(NAs)}$) & unrelaxed & relaxed & As-for-N \\
\hline
$3^3$ & \multirow{2}{*}{$1.44$} & \multirow{2}{*}{1.61} & 1.19 & 0.58 & 1.39 \\
$6^3$ &                         &                       & 1.39 & 1.06 & 1.51 \\
\end{tabular}
\end{ruledtabular}
\end{table}

\subsection{Extension of N states in real space}
\begin{figure}
 \includegraphics[width=.5\textwidth]{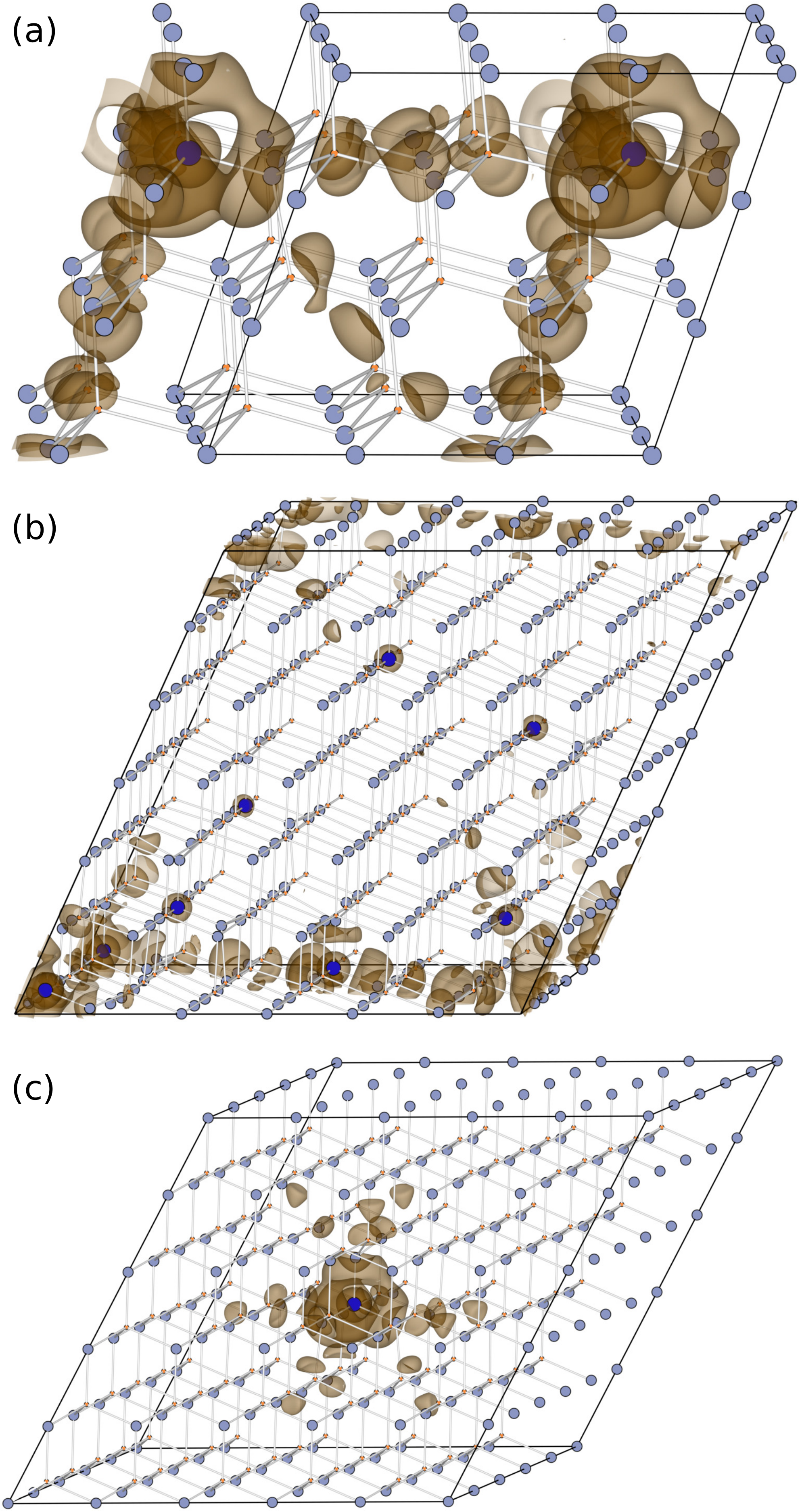}
 \caption{Partial charge density of the first conduction band of Ga(NAs) with $3.7\,\%$ N (a) $3^3$ supercell, b) $6^3$ supercell) and $0.8\,\%$ N (c) $5^3$ supercell). Isosurface at $10\,\%$ of the respective maximal value. Ga: light blue, As: orange, N: blue; Ga and As atoms are shown in smaller size.}\label{fig:ganas-chgdens}
 \end{figure}

The partial charge densities of the first conduction band of both cells discussed in the previous section are shown in FIG.~\ref{fig:ganas-chgdens}.
As was already described previously, \cite{Kent2001} for the too small $3^3$ supercell the N-atom interacts with its own translational image (FIG.~\ref{fig:ganas-chgdens}a).
Also, the charge density spreads along the \{110\}-zigzag chains.
In the case of the larger $6^3$ supercell, the partial charge density is spread mainly over three N-atoms and the chains in between those (FIG.~\ref{fig:ganas-chgdens}b).
While some artificial periodicity clearly remains, there is no such spurious interaction between translational images as for the smaller cell.
For an unrelaxed lattice, the N state is localized at the N atom itself, without spreading through the lattice (not shown).
Thus, the strong combined relaxation effect for too small supercells described above is well reflected in the real space electronic structure.
The corresponding partial charge density for the $5^3$ supercell used to study structural effects is shown as well (FIG.~\ref{fig:ganas-chgdens}c).
It is sufficiently localized at the N-atom and contained within the unit cell to avoid artificial interaction over the cell boundaries.
These observations agree well with the structural finding above, since the CB states are extended mainly along the same zigzag lines, on which the strain field propagates.
In a real space picture, the reduced interatomic distance along these lines increases the orbital overlap, thereby increasing the extension of the N-centered state.

\subsection{Band gaps of dilute nitrides for concentrations up to 11\%}
While the $6^3$ supercell clearly is better suited than the smaller $3^3$ supercell, it remains to be determined if it is large enough to reproduce accurate band gaps for nitrides over a relevant concentration range.
To this end, both simple supercells and SQS cells of increasing size are compared for various concentrations with reference values from an established CBAC\cite{Vurgaftman2003} model in TABLE~\ref{tab:gap-compare}.
For all concentrations, the $3^3$ supercells perform not only considerably worse than the $6^3$ cells, but show deficits for increasing concentrations of N, especially for specific arrangements of atoms in the cell (a particular arrangement of three N atoms bound to one Ga atom almost closes the band gap for $11.1\,\%$ N).
An average deviation from the reference of $-0.50\,\text{eV}$ demonstrates the effect of the spurious interaction between electronic states of periodic images of N atoms for the $3^3$ supercells.
The larger supercells on the other side agree with the reference values within the accuracy that can be expected from the functional, on average underestimating the reference by approximately $0.06\,\text{eV}$, and are thus large enough to overcome the artificial band gap reduction caused by the periodic boundary conditions.
This shows, that the combination of an accurate, yet feasible functional with moderately large unit cells allows quantitative, predictive computations of dilute nitride band gaps.

Furthermore, SQS cells were investigated regarding their suitability for predictive band gap computations.
This approach is designed to efficiently model random structures with periodic boundary conditions.
As such it saves the need for sampling and generally reduces the required cell size.
Given the notorious difficulty in the prediction of dilute nitride band gaps, however, a control is indicated.
The SQS cells of different sizes yield a range of band gaps that are generally close to one another for a given concentration.
For the lower concentrations of $3.7\,\%$ and $7.4\,\%$~N, the tested SQS cells agree well with the reference and clearly outperform the $3^3$ supercells.
This is especially noteworthy for the SQS-54 of the same size.
For $7.4\,\%$~N in particular, the agreement to the reference is remarkable for the SQS-54 ($+0.02\,\mathrm{eV}$) and the SQS-216 ($+0.01\,\mathrm{eV}$).
For $11.1\,\%$ N the picture changes somewhat, as the SQS cells strongly underestimate the band gap compared to the reference, yielding only slightly higher band gaps than the $3^3$ supercells.
This holds for all tested SQS cells, thus increasing the size does not mitigate this effect.

Based on the observations described above, it appears that the unavoidable periodicity prevents the strain field and the N-centered CB states from decaying sufficiently for high concentrations such as $11.1\,\%$ N even for larger SQS cells.
The $6^3$ supercells on the other hand seems to fulfill this requirement even for larger concentrations.
However, for $N^3$ supercells, scattering of band gap values is inevitable, making sampling a necessity.
Thus, for smaller concentrations of N in Ga(NAs), SQS cells remain a viable, very efficient alternative, especially since small SQS cells are sufficient in the cases where the SQS approach is applicable.

\begin{table}
\caption{Band gaps of Ga(NAs) with various N-concentrations and supercell types and sizes. Cubic $3^3$ and $6^3$ supercells as well as SQS-54, -108, -162, and -216 cells have been used for N concentrations of $3.7$, $7.4$, and $11.1\,\%$. Values were obtained by averaging over several arrangements for the $N^3$ supercells with the exception of the $3^3$ cell for $3.7\,\%$, for which only one arrangement exists. Direct band gaps in eV.}\label{tab:gap-compare}
\begin{ruledtabular}
\begin{tabular}{rrrrrrrr}
{\% N} & $3^3$ & $6^3$ & {S-54} & {S-108} & {S-162} & {S-216} & {Ref.\cite{Vurgaftman2003}} \\
\hline
 3.7 & 0.58 & 1.00 & 0.91 & 0.99 & 1.30 & 1.17 & 1.06 \\
 7.4 & 0.31 & 0.77 & 0.87 & 0.92 & 0.93 & 0.86 & 0.85 \\
11.1 & 0.20 & 0.63 & 0.25 & 0.35 & 0.28 & 0.24 & 0.68 \\
RMSD & 0.50 & 0.06 & 0.26 & 0.20 & 0.27 & 0.26 & \\
\end{tabular}
\end{ruledtabular}
\end{table}

\begin{figure}
 \includegraphics[width=.8\textwidth]{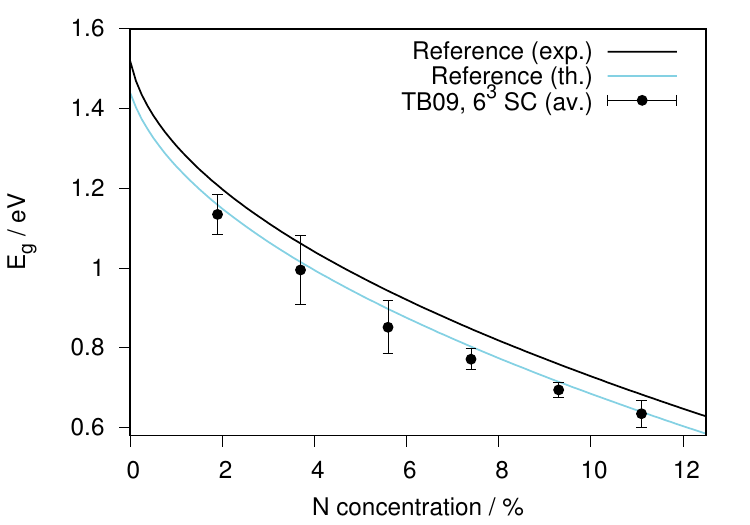}
 \caption{Band gaps of Ga(NAs) for a $6^3$ supercell averaged over four arrangements each for the three concentrations up to $5.6\,\%$ N, three arrangements each for $7.4$ and $9.3\,\%$ N and two for $11.1\,\%$. The references are based on the CBAC model parameters as given by Ref.~\cite{Vurgaftman2003} with the experimentally derived band gap (exp.) and with the TB09 band gap (th.) of GaAs (see TABLE~\ref{tab:functionals}).}
 \label{fig:gaps-sc6}
 \end{figure}

FIG.~\ref{fig:gaps-sc6} shows band gaps we calculated using the $6^3$ supercell and the band gap decrease as predicted by the CBAC model. For the three lower concentrations ($1.9\,\%$, $3.4\,\%$, and $5.6\,\%$), four arrangements have been averaged, three for $7.4\,\%$ and $9.3\,\%$ and two for $11.1\,\%$.
Since different local motifs are more likely to be repeated for higher concentrations, the scattering of values tends to decrease (indicated as error bars in FIG.~\ref{fig:gaps-sc6}).
Using the experimental band gap for GaAs $E_g = 1.52\,$eV and the CBAC model parameters from Ref.~\cite{Vurgaftman2003}, the predicted band gaps are overall larger than our calculated ones.
However, since our supercell calculations are based on the TB09 functional, which yielded a band gap of $E_g = 1.44\,$eV (see TABLE~\ref{tab:functionals}), this theoretical (th.) value should be used with the CBAC model in order to allow for comparability.
And indeed, an excellent agreement between CBAC (th.) and the supercell band gaps is found in FIG.~\ref{fig:gaps-sc6}.

It can be concluded, that a $6^3$ supercell is indeed large enough to counteract the artificial interaction with periodic images that plagues smaller supercells, while SQS of smaller size can be used for computing band gaps at lower N-concentrations, yet they seem too small at higher concentration.

\subsection{Effective band structure of Ga(NAs)}
\begin{figure}
 \includegraphics[width=.6\textwidth]{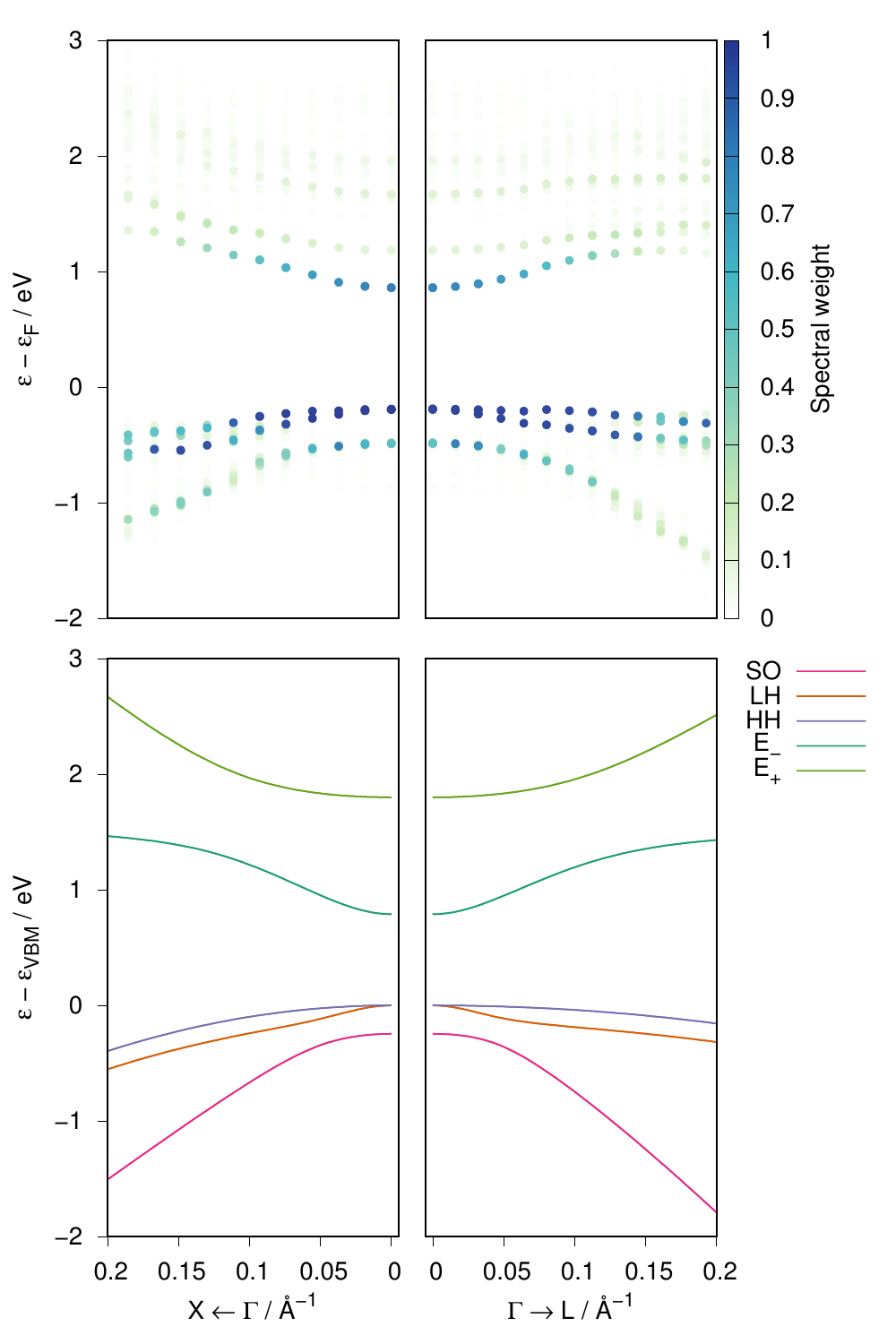}
 \caption{Effective band structure of a $6^3$ supercell of Ga(N$_{0.037}$As$_{0.963}$) along $\Delta$ and $\Lambda$ high symmetry axes (top row) contrasted to a $\mathbf{k}\cdot\mathbf{p}$-band structure obtained from a 10 band conduction band anti-crossing Hamiltonian (bottom row).}
 \label{fig:ebs-sc6}
 \end{figure}

In order to predict optical properties from \textit{ab initio} calculations, it is necessary to go beyond the computation of band gaps alone.
To explore the possibility for the system at hand, a part of the band structure along the $\Delta$ and $\Lambda$ high symmetry axes in the immediate vicinity of the $\Gamma$-point was computed and unfolded using a custom unfolding routine for the single arrangement of $\mathrm{Ga}_{216}\mathrm{N}_8\mathrm{As}_{208}$ already used above.
The effective band structure (EBS) is shown and compared with a 10-band CBAC $\mathbf{k}\cdot\mathbf{p}$-band structure\cite{Vurgaftman2003,Shan1999,Lindsay1999,OReilly2002} in FIG.~\ref{fig:ebs-sc6}.
Despite the simplistic CBAC model with a single nitrogen level, the overall agreement between both approaches is rather good, especially for the valence bands.
The two conduction bands described by the CBAC-model can be identified in the DFT derived EBS, however, one additional band with a spectral weight of roughly $0.2$ and several other bands with lower spectral weight occur in the EBS.
The $\mathrm{E}_{+}$-band of the $\mathbf{k}\cdot\mathbf{p}$-band structure can be considered as an average of the higher bands in this context.
However, the additional band right above the lowest conduction band is not accounted for by the CBAC-model.
Since the $\mathrm{E}_{-}$-band has a higher spectral weight and is lower in energy, it is the most relevant for the computation of optical properties.
Thus, this issue does not invalidate the CBAC-$\mathbf{k}\cdot\mathbf{p}$-model as a starting point for this purpose.
Nevertheless, the existence of an intermediate band may be relevant for higher excitations.

\section{Conclusions and Outlook}
We revisited the long standing issue of \textit{ab initio} calculations of the electronic structure of dilute nitrides by combining a density functional producing quantitative band gaps (TB09) with a converged supercell approach.
Our main finding is, that with this approach band gaps in excellent agreement with the established CBAC model can be achieved for the Ga(NAs) system using feasible supercell sizes.
The required supercell size is determined by the extension of the distortion field the N atom introduces in the GaAs lattice and its electronic states.
Even with a functional that produces good band gaps by itself, this criterion must be met to avoid spurious interactions of the N atom with its translational image, which manifests itself in the strong overestimation of relaxation effects on the band gap.
Comparison with the SQS-approach showed, that this design principle allows us to reach superior results compared to simple supercells of the same size also for this material system.
At higher concentrations, however, simple supercells of sufficient size allow for an easy and fast access to the electronic properties of dilute nitrides and allow to systematically study the effect of N atom arrangements on these.
Building on the results presented in this paper, it is now possible to obtain further data on the electronic structure of dilute nitrides, especially predict the properties of new, unknown materials, by computing effective band structures and extracting carrier effective masses and other band structure parameters.

\begin{acknowledgments}
This work was supported by the DFG in the framework of the Research Training Group ``Functionalization of Semiconductors'' (GRK~1782). The authors thank the HRZ Marburg, CSC-LOEWE Frankfurt and HLR Stuttgart for providing computational resources.
\end{acknowledgments}

\bibliography{literature}
\end{document}